\documentclass[lettersize,journal]{IEEEtran}

\usepackage[english]{babel}
\usepackage{amsmath,amsfonts}
\usepackage{array}
\usepackage[caption=false,font=normalsize,labelfont=sf,textfont=sf]{subfig}
\usepackage{textcomp}
\usepackage{stfloats}
\usepackage{url}
\usepackage{verbatim}
\usepackage{graphicx}
\usepackage{cite}
\usepackage{amssymb}
\usepackage{amsbsy}
\usepackage{amsmath}
\usepackage{amsthm} 
\usepackage{setspace}
\usepackage{gensymb} 
\usepackage{amsmath}
\usepackage{gensymb}
\usepackage{blindtext}
\usepackage{enumitem}
\usepackage{xcolor}
\usepackage{amssymb}
\usepackage{amsbsy}
\usepackage{amsmath}
\usepackage{setspace}
\usepackage{gensymb} 
\usepackage{amsmath}
\usepackage{gensymb}
\usepackage{blindtext}
\usepackage{enumitem}
\usepackage{xcolor}
\usepackage{amsfonts}
\usepackage{mathrsfs}
\usepackage{amssymb}
\usepackage{algorithm}
\usepackage{algpseudocode}

\usepackage{amsmath}
\usepackage{graphicx}

\begin{document}
\title{Learn to Adapt to New Environment from Past Experience and Few Pilot}
\author{Ouya Wang, Jiabao Gao, and Geoffrey Ye Li
\thanks{Ouya Wang and Geoffrey Ye Li are with the Department of Electrical and Electronic Engineering, Faculty of Engineering, Imperial College London, London SW7 2BX, U.K. (e-mail: ouya.wang20@imperial.ac.uk; geoffrey.li@imperial.ac.uk).}%
\thanks{Jiabao Gao is with the Institute of Information and Communication Engineering, Zhejiang University, Hangzhou 310027, China. (e-mail: gao\_jiabao@zju.edu.cn)}}
\maketitle

\begin{abstract}
In recent years, deep learning has been widely applied in communications and achieved remarkable performance improvement. Most of the existing works are based on data-driven deep learning, which requires a significant amount of training data for the communication model to adapt to new environments and results in huge computing resources for collecting data and retraining the model. In this paper, we will significantly reduce the required amount of training data for new environments by leveraging the learning experience from the known environments. Therefore, we introduce few-shot learning to enable the communication model to generalize to new environments, which is realized by an attention-based method. With the attention network embedded into the deep learning-based communication model, environments with different power delay profiles can be learnt together in the training process, which is called the learning experience. By exploiting the learning experience, the communication model only requires few pilot blocks to perform well in the new environment. Through an example of deep-learning-based channel estimation, we demonstrate that this novel design method achieves better performance than the existing data-driven approach designed for few-shot learning.
\end{abstract}


\section{Introduction}

Deep learning (DL) can address the intricate correlation among variables, especially those that are difficult to accurately describe with mathematical models \cite{goodfellow2016deep}, which allows us to design wireless communication systems without requiring expert knowledge. Therefore, it has been used to develop communication systems and received widespread attention for its effectiveness \cite{ye2021deep}. For the data-driven method in \cite{ye2017power}, a deep neural network (DNN) is adopted to replace the channel estimator and the signal detector in the orthogonal frequency division multiplexing (OFDM) receiver. The end-to-end design in \cite{ye2021deep} and \cite{ye2020deep} use two DNNs representing the transmitter and receiver, respectively. Recently, the spirit of the end-to-end model has been extended to semantic communication \cite{xie2021deep}. These DL-based wireless systems have demonstrated impressive performance in the additive white Gaussian noise (AWGN) channel \cite{ye2017power} and frequency-selective channels in \cite{ye2021deep}. However, only single communication environment is considered in \cite{ye2021deep} and \cite{ye2017power}. It is a challenge for the DL-based communication system to be adapted to new environments with different power delay profiles (PDPs) or distortions. 

DL has been highly successful in data-intensive applications but is often hampered by a small available training dataset \cite{wang2020generalizing}. The DL-based channel estimation (CE) is often purely data-driven. For any particular channel propagation environment, a large number of pilot blocks or labelled data are required for the training, which is usually performed offline.

Using a small dataset to train a DNN with a large number of parameters can easily lead to over-fitting \cite{goodfellow2016deep}. Few-shot learning (FSL) has been proposed to tackle that issue. Using prior knowledge, it is possible to quickly generalize a model to a new task with only a few available samples \cite{wang2020generalizing}. Meta-learning has been the most common framework for FSL in recent years. In \cite{park2020meta}, common initialization parameters that enable fast training on any channel have been found using the meta-learning approach. From \cite{park2020meta}, significant training speed improvement and an efficient communication model can be obtained only through one iteration of gradient descent. Accretionary learning \cite{wei2021accretionary} is designed to accumulate learned knowledge and acquire new knowledge. The idea of accretionary learning can be applied to achieve the goal of FSL, where knowledge is learned and accumulated independently during offline training. For accretionary learning, few new parameters are trained and combined with learned knowledge to acquire new knowledge online. In \cite{jiang2021ai}, an online training system, called SwitchNet, can capture the features of the new propagation environment. In SwitchNet, multiple DNNs are pre-trained, each for a specific propagation environment. Only a small dataset is required to linear combine outputs of those DNNs in online adaption. However, the propagation environments tested for this method are limited in \cite{jiang2021ai}. In addition to meta-learning and accretionary learning that can be potentially used to realize FSL, model-driven methods also require less data for training due to fewer trainable parameters. In \cite{jin2021adaptive}, the adaptivity of the channel estimator is enhanced by designing a hypernet to generate parameters for the model-driven based wideband mmwave system. However, for the data-driven CE, the parameter set is enormous, which is challenging for the hypernet to generate all parameters.

In this paper, we focus on a data-driven CE system that can be quickly adapted to a new environment. In Section II, we introduce the existing DL-based approach for CE and briefly review the attention mechanism. Section III formulates the problem and introduces the mechanism for attention generators. In Section IV, we present the FSL method. In Section V, we compare our FSL method with other related ones. Section VI concludes the paper and discusses the potential of developing FSL with some other techniques. The main contributions of this paper are as follows:
\begin{description}[font=$\bullet$~\normalfont\scshape\color{black}]
\item We propose to use the attention mechanism for the CE model to realize FSL, where attention networks generate weights for each feature vector under multiple rules for dynamic adjustment. To the best of our knowledge, this is the first work to introduce the attention mechanism in wireless communications to realize FSL.

\item We design a task-attention model to enhance the generalization ability for various distributed training data and improve testing performance in the new environment. By using few pilot blocks in the new environment, the task-attention network adds the knowledge of the new environment to feature maps by generating attention vectors in the channel domain \footnote{Channel domain mentioned in this paper is a term of the convolutional neural network, which represents a dimension of the feature map, rather than the communication model.}. 

\item We introduce the cross-attention mechanism to find the correlation between the support and the query blocks in the spatial domain, which leads to a higher estimation accuracy in initialization blocks. This is the first work applying the cross-attention mechanism to wireless communication problems.

\item The cross-attention model embedded initialization network is proposed to produce the input of the CE backbone. The query block is firstly initialized according to the efficient feature embedding from support blocks and then sent into the CE backbone. We develop a method that takes the most advantage of the support blocks to improve CE for the query block. 
 
\end{description}

\section{Related Works}

In this section, we briefly review recent advances in DL-based CE and introduce the working mechanism of SwitchNet, which is closely related to our work. Furthermore, the attention mechanism and its applications in wireless communication are described briefly.

\subsection{Deep Learning in CE}
DL has emerged as an effective tool for CE in wireless systems. A new DL-based receiver design, ComNet, has been proposed for an OFDM system to deal with frequency-selective fading channels \cite{gao2018comnet}, where two cascaded DNNs are utilized for the CE and signal detection (SD), respectively. For ComNet, the estimated channel is used to recover the transmitted data. With the expert knowledge embedded, ComNet is more predictable and explainable than the fully-connected DNN-based receiver in \cite{ye2017power}, which treats the joint channel estimator and signal detector as a black box. Due to the strong fitting ability of DNN and the end-to-end training mechanism, the SD network can still recover the transmitted signal even if the output of the CE network is far away from the real channel. Therefore, the output of CE can be regarded as a feature representation rather than accurate instantaneous channel coefficients.

\begin{figure*}[h!]
\centering
\includegraphics[width=0.8\textwidth]{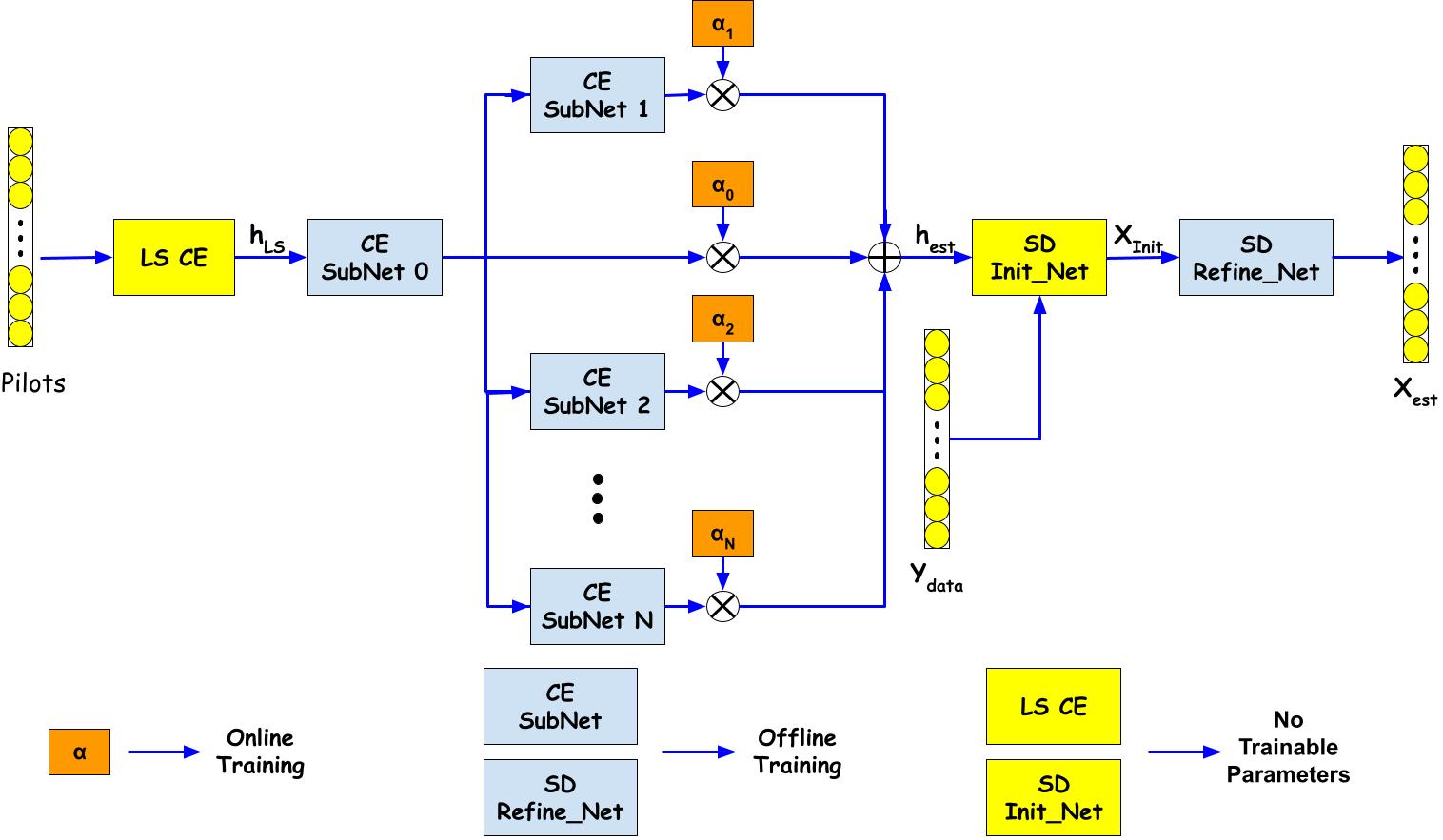}
\caption{The structure of the attention mechanism embedded into the end-to-end network. The figure only demonstrate the attention network embedded into the first two convolutional layers for the whole system.}
\label{fig:switchnet}
\end{figure*}

Recently, SwitchNet has been proposed to provide a more accurate CE and enable the system to adapt quickly to the new environment. The architecture of the SwitchNet is shown in Figure~\ref{fig:switchnet}. It consists of CE and SD and is similar to the conventional receiver. In \cite{jiang2021ai}, the structure of the CE network is designed for online adaption. The CE network consists of least-square (LS) CE and five CE SubNets, which are all implemented by neural networks. The CE outputs are linearly combined with parameters $\boldsymbol{\alpha}$. CE SubNet 0 performs the basic CE. Multiple CE SubNets are used as compensating networks for CE SubNet 0 to adapt propagating environments in the training set. Each compensation network aims at a specific propagation environment. During online adaption, each compensation network is governed by a trainable parameter $\alpha_i$ that controls its contribution to fit the new environment. Since there are only few trainable variables during testing for the new environment, only a small batch of samples is required for the adaption. Denote $\textbf{W}_i$ and $\boldsymbol{\theta}_i$ as the multiplicative parameter weights and bias, respectively. The estimated channel can be expressed as,

\begin{align}
\boldsymbol{h}_{est} &= (\sum_{i=1}^M \alpha_i \boldsymbol{W}_i +\alpha_0 \boldsymbol{I})(\boldsymbol{W}_0 \boldsymbol{h}_{ls} +\boldsymbol{\theta}_0) + 
\sum_{i=1}^M\alpha_i \boldsymbol{\theta}_i, 
\end{align}

\noindent where $\boldsymbol{h}_{ls}$ is channel coefficients calculated through LS. After more accurate channel coefficients $\textbf{h}_{est}$ are estimated, the signal detection (SD) network is employed to recover the transmitted data. In order to compare the CE performance with our proposed method, only the CE part of SwitchNet will be used sometime. The above is the offline process. The online adaption aims at learning a combination of the CE SubNets using the support blocks, where true channel coefficients are known for the SwitchNet support blocks. The online fine-tuning process is to learn a set of $\boldsymbol{\alpha}$ to minimize the mean-squared error between the true channel coefficients and the output of CE SubNet, $\textbf{h}_{est}$. Such an online adaption mechanism allows the model to be adapted to different propagation environments and makes the system more robust than conventional DL-based communication systems.

\subsection{Attention Mechanism}

Convolutional Neural Networks (CNN) performs classification or regression tasks by generating feature maps for samples. Traditional CNN often treats the extracted features equally, limiting the generalization of a model in some cases. When the category of testing samples is highly separate data, like instantaneous channel coefficients from different environments, the importance of these extracted features is different. For example, the features for coping with channel distribution within a specific angular region might have limited even with negative impacts on estimating channels from another region or environment. Therefore, the attention mechanism is employed to generate weights for each extracted feature vector so that more critical features have more significant contributions \cite{gao2021attention}.

The theory of attention in DL has been first proposed in \cite{mnih2014recurrent}, which adopts recurrent neural network (RNN) and reinforcement learning (RL) to obtain attention in the spatial domain. Then attention mechanism becomes popular with the introduction of squeeze-and-excitation network (SENet) \cite{hu2018squeeze} and multi-head attention \cite{vaswani2017attention}. SENet is designed to recalibrate channels by using attention weights adaptively. It advances computer vision. Our proposed attention approach is developed based on this theory. We use self-attention to represent the working mechanism of SENet subsequently.

Attention mechanism has been employed to help the DL-based communication system, such as channel state information (CSI) compression \cite{cai2019attention}, channel compression\cite{xu2021wireless} and CE \cite{luan2022attention} \cite{gao2021attention}. It can enhance the estimation accuracy for channels with highly separate distributions in \cite{gao2021attention}. Therefore, it has the potential to improve the learning capacity of the DL model by accumulating more experience learning from datasets with different distributions.

\section{Problem Formulation}
In this section, we will introduce our FSL problem scenario and provide necessary information about cross attention and task attention in FSL application.

\subsection{Problem Scenario}
The mobile communication channel in our problem scenario is multipath propagation, which leads to serving dispersion of the transmitted signal. The PDP indicates the distribution of transmitted power over various paths in propagation \cite{saha2016power}. The channel impulse response (CIR) in a multipath propagation channel can be written as,
\begin{align}
    \boldsymbol{h}(t) = \sum_{l=1}^LA_le^{-j\phi_l}\delta(t-\tau_l),
\end{align}
\noindent where $L$ is the number of resolvable paths. $A_l$, $\phi_l$ and $\tau_l$, represent attenuation, phase shifts, and delay of time arrival in the $l^{th}$ path. The channel PDP is calculated from the spatial average of $\left| \boldsymbol{h}(t) \right|^2$ over a local area and represents small-scale multipath channel statistics. In this paper, we define the propagation environment as an area shared with the same PDP. We consider channels with fast Doppler shifts, where instantaneous channel coefficients vary quickly.     

We assume that the channel estimator is designed and trained offline for several propagation environments. Then we investigate how we can use the previous experience together with a few pilot blocks to estimate the channel in a new environment. In the beginning, all symbols in the blocks are pilots in the new environment, which are referred to as support blocks $\mathcal{T}_{support}$. We can learn the features of the new environment well from these pilot blocks, which can be done when mobile devices are on standby. Then the pilot portion in each block is decreased so that transmission becomes more efficient. Such blocks with fewer pilots are called query blocks $\mathcal{S}_{query}$. Due to fast Doppler shifts, $\mathcal{T}_{support}$ and $\mathcal{S}_{query}$ may differ more significantly compared with slow time-varying channel cases. However, they still share environmental features, such as the PDP. 

\begin{figure*}[h!]
\centering
\includegraphics[width=0.75\textwidth]{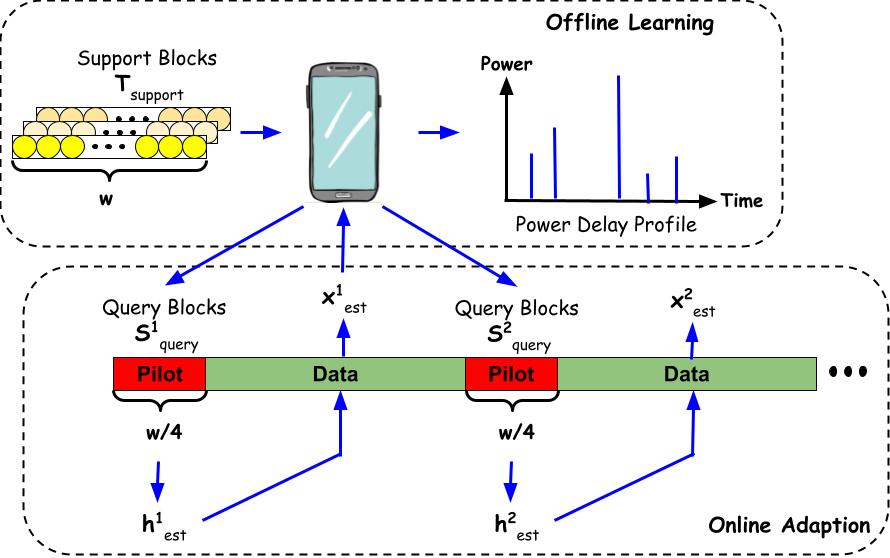}
\caption{An overview of how the communication system adapts to the new environment with few pilot blocks}
\label{fig:Support_Query_Process}
\end{figure*}
As shown in Figure~\ref{fig:Support_Query_Process}, each support block, $\mathcal{T}_{support}$, includes whole $w$ pilots in an OFDM block while the query block $\mathcal{S}_{query}$ only contains $\frac{w}{4}$ pilots. The communication system will learn environmental features offline from $\mathcal{T}_{support}$. In the online adaption, these features are employed to enhance the CE from $\mathcal{S}_{query}$. The communication system can estimate channel $\textbf{h}^i_{est}$ from the $i^{th}$ query block and then recover the transmitted data $\textbf{x}^i_{est}$ in the $i^{th}$ data block. The accuracy of CE plays a decisive role in transmitted data detection. Therefore, we develop an FSL approach to quickly adapt the communication system to the new environment. With this FSL approach, we can still achieve good CE results for query blocks in the new environment with the guidance of few support blocks.

Support blocks and query blocks in our problem are different from those in conventional FSL problems. In conventional settings, the support dataset comes with true labels or values, while the query dataset is unlabelled. In our cases, the support blocks, $\mathcal{T}_{support}$, have no corresponding true values since it is hard to obtain the accurate channel coefficients in real communication systems due to various inherent uncertainties of the wireless channel. We only use channel coefficients estimated by LS from the pilot blocks. Different from SwitchNet adaption \cite{jiang2021ai}, where online fine-tuning is necessary and the true channel coefficient requires to be known during fine-tuning, $\mathcal{T}_{support}$ is utilized directly to help the DL-based CE model generalization without any fine-tuning. Support blocks $\mathcal{T}_{support}$ have more pilots compared with query blocks, which means the channel estimated from $\mathcal{T}_{support}$ has higher accuracy. Therefore,  $\mathcal{T}_{support}$ contains more channel information and more accurate environment features can be extracted compared to that included in $\mathcal{S}_{query}$. 
Since instantaneous channels from the same propagation environment share many features, the environment features extracted from $\mathcal{T}_{support}$ can be employed to enhance the CE for $\mathcal{S}_{query}$.

The essence of our FSL problem is to use a small number of pilot blocks with more information ($\mathcal{T}_{support}$) to guide blocks with less channel information ($\mathcal{S}_{query}$) for efficient CE in new environments. With only a small number of support blocks, the estimation accuracy of query blocks can be close to the testing accuracy boundary, which is obtained by testing the query blocks with the model trained with sufficient pilot blocks of the same environment. 
The approach to achieve the goal of FSL is to find out useful features through a small number of support blocks and then use these features to improve the accuracy of the channel estimated from the query block.

The attention mechanism is employed to realize the approach. The task attention uses global environment features extracted from all support blocks to give dynamic adjustment in the process of estimating channels from $\mathcal{S}_{query}$. In comparison, the cross-attention uses local feature correlation between each support and query block to enhance channel feature embedding. For each $\mathcal{S}_{query}$, we will select different channel features from $\mathcal{T}_{support}$ and embed them in $\mathcal{S}_ {query}$ to improve CE. This process is called CE guidance. Such guidance depends on the individual pilot block, while the adjustment from task attention is only related to the environment. Both attention approaches have meta-learner to guarantee effectiveness when deployed to new environments. These two attention mechanisms will be introduced subsequently.  
\begin{figure*}[!htb]
\centering
\subfloat[The structure of the cross-attention network][The structure of the cross-attention network]{
\includegraphics[width=0.5\textwidth]{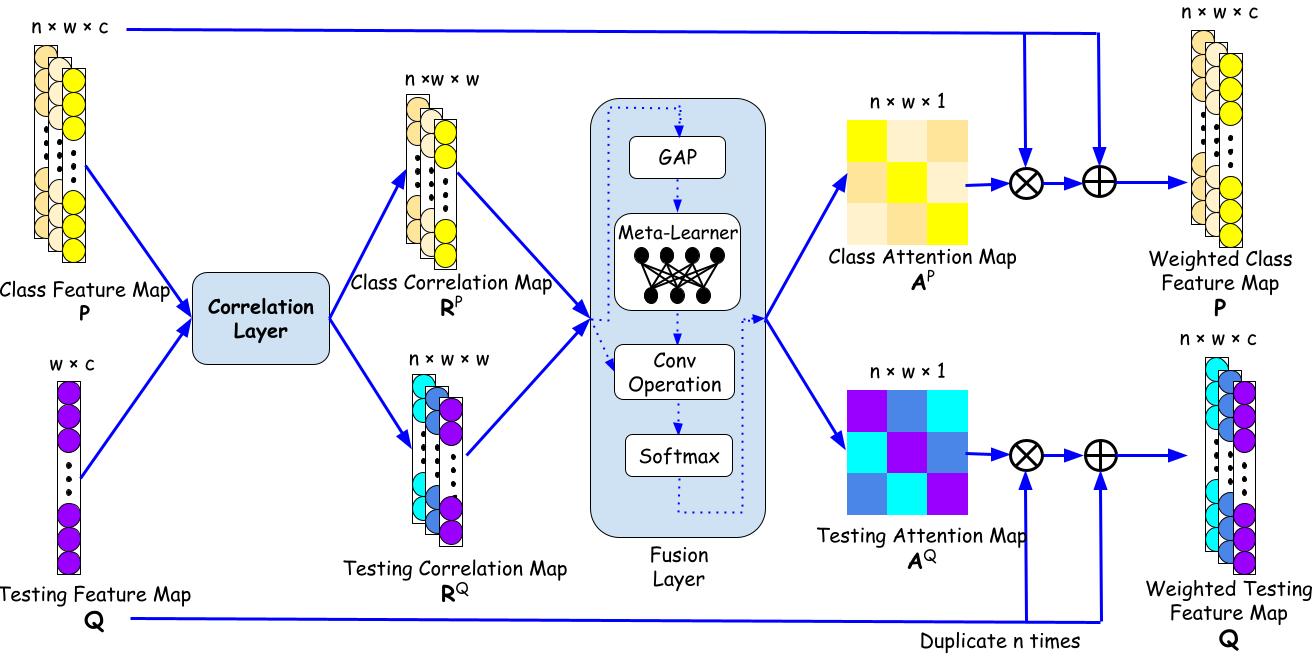}
\label{fig:subfig11}}
\subfloat[The structure of the task-attention network][The structure of the task-attention network]{
\includegraphics[width=0.5\textwidth]{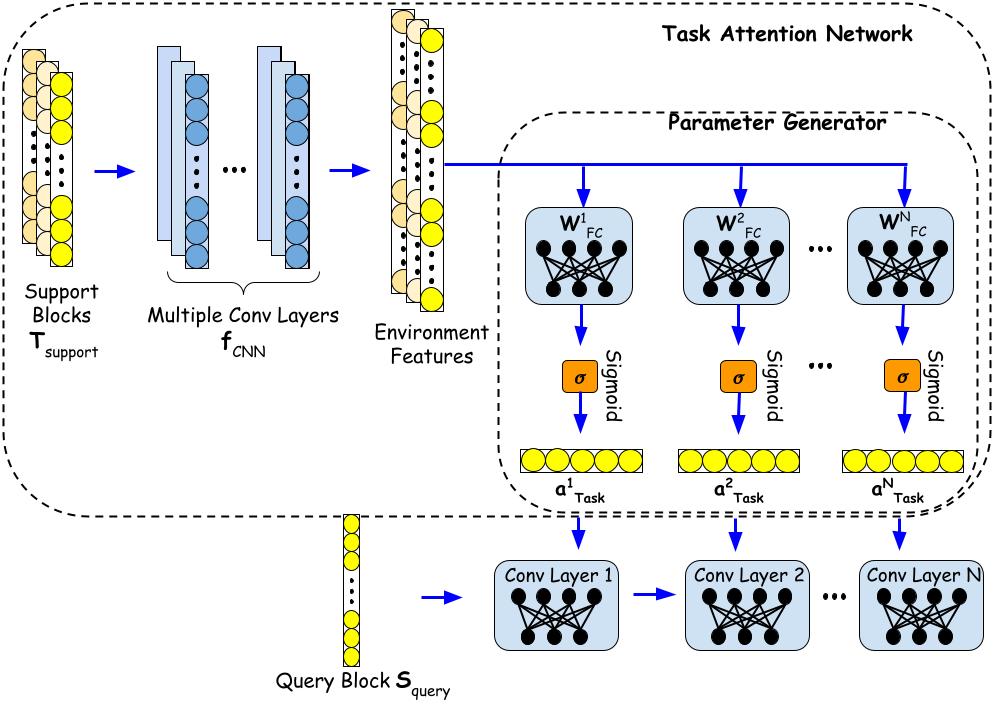}
\label{fig:subfig22}}
\caption{The structure of two attention networks}\label{fig:self&task}
\end{figure*}

\subsection{Cross-Attention}
Conventional attention models (e.g. SENet) are not effective for FSL since they usually find out important features of the test samples only based on the prior of the training dataset and result in poor generalization performance to unseen samples. Furthermore, a large number of samples are required to learn the attention mechanism for any specific class. Instead of using each sample's own feature map to draw attention, the cross-attention model (CAM), proposed in \cite{hou2019cross}, employs semantic relevance between support and query set features to generate attention maps in pairs. The query set contains the unlabeled samples from unseen classes, while the support set has few labelled samples from the corresponding classes. With such an approach, only few labelled samples are required to highlight the important regions so that more discriminative features can be extracted.

The CAM is illustrated in Figure~\ref{fig:self&task} (a) in our problem scenario. The class feature map, $\textbf{P}\in \mathbb{R}^{n \times w \times c}$, is extracted from the support blocks and the testing feature map, $\textbf{Q}\in \mathbb{R}^{w \times c}$, is extracted from the query block, where $c$ and $w$ refer to the number of channels and the width of the feature map generated from one block, $n$ represents the number of support blocks. For each testing feature map, there are $n$ class feature maps to help extract attention vectors. CAM will produce cross attention weights $\textbf{A}^{p}$ ($\textbf{A}^{q}$) in the spatial domain for the feature map $\textbf{P}$ ($\textbf{Q}$). The testing feature map computes the cross-attention with each class feature map. A total of $n$ pairs of attention maps will be generated. The testing feature map, $\textbf{Q}$, is duplicated $n$ times in order to weight with different attention maps.

First, the correlation map between feature maps $\textbf{P}$ and $\textbf{Q}$ is calculated for the generation of the cross-attention map. Such a correlation map $\textbf{R}$ refers to the semantic relevance between $\textbf{p}_i$ and $\textbf{q}_i$ with cosine distance, where $\textbf{p}_i$ and $\textbf{q}_i \in \mathbb{R}^{c}$ are the $i^{th}$ spatial position in $\textbf{P}$ and $\textbf{Q}$, respectively. The correlation layer computes the semantic relevance between $\textbf{p}_i$ and $\textbf{q}_j$ to get the correlation map $\boldsymbol{R} \in \mathbb{R}^{nw \times w}$ as 
\begin{align}
    \boldsymbol{R}_{ij} = (\frac{\boldsymbol{p}_i}{\|\boldsymbol{p}_i\|_2})^T(\frac{\boldsymbol{q}_j}{\|\boldsymbol{q}_j\|_2}), i = 1,...,nw, j = 1,...,w. 
\end{align}

There are two correlation maps, $\textbf{R}^P = [\textbf{r}^p_{1},\textbf{r}^p_{2} ... \textbf{r}^p_{nw}]^T$ and $\textbf{R}^Q = [\textbf{r}^q_{1},\textbf{r}^q_{2} ... \textbf{r}^q_{w}]$, where $\textbf{r}^p_{i} \in \mathbb{R}^{w}$ denotes the correlation between the local support block's feature vector $\textbf{p}_i$ and all query blocks feature vectors $\{\textbf{q}_j\}^{w}_{j=1}$, and $\textbf{r}^q_{j} \in \mathbb{R}^{nw}$ is the correlation between the query block feature vector, $\textbf{q}_j$, and all support block feature vectors $\{\textbf{p}_i\}^{nw}_{i=1}$. The correlation vector $\textbf{r}^p_{i}$ is formulated as:
\begin{align}
    \boldsymbol{r}^p_{i} = \boldsymbol{p}^T_i\boldsymbol{Q}^T, i = 1,...,nw. 
\end{align}

Then, a fusion layer is employed to generate attention maps based on $\textbf{R}^p$ and $\textbf{R}^q$. The fusion layer applies global average pooling (GAP) to the correlation map and sends it into a meta-leaner to generate kernel $\textbf{w}_{kernel}$ in the spatial domain. The convolutional operation is taken between the correlation map $\textbf{R}^p$ and kernel $\textbf{w}_{kernel}$, which aims at fusing each local correlation vector $\{\textbf{r}^p_i\}^{nw}_{i=1}$ of $\textbf{R}^p$ into an attention value. Finally, softmax is employed to normalize the attention value. The attention value at the $i^{th}$ place can be expressed as
\begin{align}
    \boldsymbol{A}^p_i = \frac{\exp(\textbf{W}_2(\sigma\textbf{W}_1(GAP(\textbf{R}^p)))^T \boldsymbol{r}^p_i/\tau)}{\sum^{nw}_{j=1}\exp(\textbf{W}_2(\sigma\textbf{W}_1(GAP(\textbf{R}^p)))^T \boldsymbol{r}^p_j/\tau)} 
\end{align}

\noindent where $\textbf{A}^p_i$ represents the attention value at the $i^{th}$ position, $\tau$ is the temperature hyperparameter in softmax function, $\textbf{W}_1$ and $\textbf{W}_2$ are parameters of meta-learner, and $\sigma$ represents to the RELU function. The meta-learner generates the kernel, $\textbf{w}_{kernel}$, that aggregates the correlations between two feature maps $\textbf{P}$ and $\textbf{Q}$ in order to draw attention to the target objects. The testing attention map, $\textbf{A}^{q}$, can be achieved in a similar way. Finally, we weight the initial feature map, $\textbf{P}$ and $\textbf{Q}$, elementwisely by $1+\textbf{A}^{p}$ and $1+\textbf{A}^{q}$ to form a more discriminative feature maps. 

\subsection{Task-Attention}
The idea of task-attention is borrowed from the task-aware feature embedding network (TAFE-Net) in \cite{wang2019tafe}, which is proposed to handle FSL in image classification. TAFE-Net consists of a meta-learner and a prediction network backbone to do classification for unlabelled images. The meta-learner module focuses on extracting feature embedding from few labelled images for the particular task and generating task-specific weights for each layer in the prediction network. TAFE-Net has shown promising results for data efficiency improvement for FSL. In our experiment settings, we estimate channel from query block, $\mathcal{S}_{query}$, in CE backbone, which acts as a prediction model in TAFE-Net. By following the working mechanism of the meta-learner in TAFE-Net, we develop a task-attention model (TAM) to improve the estimation accuracy.              

In Figure~\ref{fig:self&task} (b), the TAM employs an extra CNN to extract features of the support blocks. Then the extracted features are sent to a single layer perceptron to generate attention vectors $\boldsymbol{a}_{task}$ for the CE backbone. The 2D convolutional layers are employed for the CNN since the input shape of the TAM is $n \times w \times k$, where $k$ equals two since each complex channel coefficient is split into two real values. The TAM in Figure~\ref{fig:self&task} (b) can be formulated in the following,
\begin{align}
    \boldsymbol{a}^i_{task} =\frac{1}{1+\exp{(\boldsymbol{W}^i_{FC}(\boldsymbol{f}_{CNN}(\mathcal{T}_{support})))}},
\end{align}

\noindent where $\boldsymbol{f}_{CNN}$ represents the multi-convolutional-layer network function, $\boldsymbol{W}^{i}_{FC}$ is the weights of the parameter generator with single fully-connected (FC) layer. The sigmoid function is utilized to limit the range of the parameters. The output attention vector for the $i^{th}$ backbone layer output is represented as $\boldsymbol{a}^i_{task}$, as shown in Figure~\ref{fig:self&task} (b).

\section{FSL for CE}
Instead of using a large amount of data, we propose an FSL approach to allow the channel estimator to adapt to the new environment with only few pilot blocks. In this section, we are going to present the attention-based CE method in detail, where the overview of the algorithm and the working mechanism of each attention model will be demonstrated.     
\subsection{Structure of FSL-based CE}

\begin{figure*}[h!]
\centering
\includegraphics[width=0.7\textwidth]{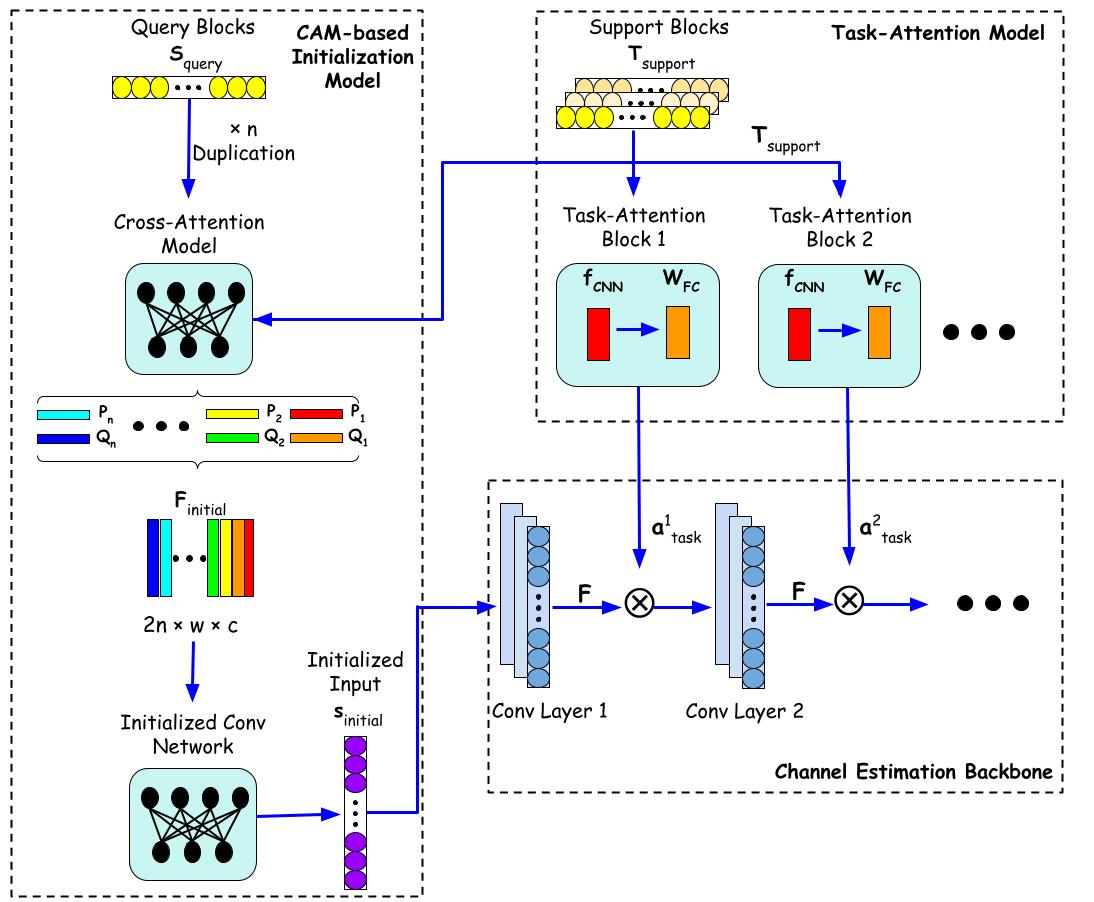}
\caption{The structure of the attention mechanism embedded into the CE backbone.}
\label{fig:end2end_attention}
\end{figure*}

As shown in Figure~\ref{fig:end2end_attention}, the FSL-based channel estimator consists of three parts: the CE backbone, the CAM-based initialization network, and the TAM. The CE backbone estimates channel coefficients from the query blocks. CNN is known for being good at exploiting correlation in the input data and is widely employed in CE applications in the frequency domain \cite{jiang2021dual} and the angular domain \cite{gao2021attention}. A multi-layer one-dimensional (1D) CNN is used for the CE backbone due to the shape of the input query blocks, as shown in Figure~\ref{fig:end2end_attention}. We design the initialization network to initialize the query blocks before sending them to the CE backbone. The initialized query blocks, $\textbf{s}_{initial}$, are used as the input of the CE backbone. The initialization network allows support blocks to provide guidance for the query block and CAM is embedded to enhance such guidance. The TAM helps the backbone to improve robustness. It selects important features in the channel domain, while the CAM helps select features for the initialization model in the spatial domain. With the cooperation of the attention mechanism and the initialization model, the CE backbone can be fast adapted to the new environment. 

We should emphasize that the above CE backbone is trained offline, corresponding to certain channel environments, such as indoor or outdoor channels. When deploying the CE backbone online, its parameters, $\theta_{CE}$, are fixed. If it is used in a new environment, the mismatch will cause significant performance degradation, which will be addressed through FSL.  

The detailed algorithm is demonstrated in Algorithm~\ref{alg1}, where $\boldsymbol{f}_{CAM}$, $\boldsymbol{f}_{CIN}$, $\boldsymbol{f}_{TAM}$, and $\boldsymbol{f}_{CE}$ refer to the functions of the CAM, convolutional initialization network, TAM and CE backbone. From  Algorithm~\ref{alg1}, the CE backbone has N+2 layers. Except for the last layer, which directly generates the CE result, the output of each other layer is multiplied by the weights, $\boldsymbol{a}_{task}$, generated by TAM. The loss function is considered in two aspects, the output of the CE backbone and the CAM-based initialization network output, both of which derive the error based on the true channel coefficients $\boldsymbol{s}_{true}$. 
\begin{algorithm}
\caption{Training the attention-based FSL for CE system} \label{alg1}
\begin{algorithmic}
\Require Learning rate $\eta$
\Require Initial Parameters $\boldsymbol{\theta}_1$ = [$\boldsymbol{\theta}_{CAM}$,$\boldsymbol{\theta}_{CIN}$]
\Require Initial Parameters $\boldsymbol{\theta}_2$ = [$\boldsymbol{\theta}_{TAM}$,$\boldsymbol{\theta}_{CE}$]
\State $\boldsymbol{g} \gets 0$
\State $\boldsymbol{\theta} = [\boldsymbol{\theta}_1, \boldsymbol{\theta}_2]$
\For{$i = 0, 1, ...$} \Comment{Loops of samples}
\State Sample a batch of $\textbf{t}$ from $\mathcal{T}_{support}$ 
\State Sample a batch of $\boldsymbol{s}$ and $\boldsymbol{s}_{true}$ from $\mathcal{S}_{query}$
\State $\boldsymbol{F}_{initial} \gets \boldsymbol{f}_{CAM}(\boldsymbol{s}, \boldsymbol{t},  \boldsymbol{\theta}_{CAM})$
\State $\boldsymbol{s}_{initial} \gets \boldsymbol{f}_{CIN}(\boldsymbol{F}_{initial}, \boldsymbol{\theta}_{CIN})$

\State $\boldsymbol{a}_{task} \gets \boldsymbol{f}_{TAM}(\boldsymbol{t}, \boldsymbol{\theta}_{TAM}^0)$
\State $\boldsymbol{F} \gets \boldsymbol{f}_{CE}(\boldsymbol{s}_{initial}, \boldsymbol{\theta}_{CE}^0)$
\State $\boldsymbol{F} \gets \boldsymbol{F} \odot \boldsymbol{a}$
\For{$j = 1, 2, ..., N$} \Comment{Loops of CE backbone layers}
\State $\boldsymbol{a}_{task} \gets \boldsymbol{f}_{TAM}(\boldsymbol{t}, \boldsymbol{\theta}_{TAM}^j)$
\State $\boldsymbol{F} \gets \boldsymbol{f}_{CE}(\boldsymbol{F}, \boldsymbol{\theta}_{CE}^j)$
\State $\boldsymbol{F} \gets \boldsymbol{F} \odot \boldsymbol{a}_{task}$
\EndFor
\State $\boldsymbol{g} \gets \bigtriangledown\mathcal{L}_{\boldsymbol{\theta}_1}(\boldsymbol{s}_{initial}, \boldsymbol{s}_{true}) +$
\Statex \qquad \qquad  $\bigtriangledown\mathcal{L}_{\boldsymbol{\theta}_2}(\boldsymbol{f}_{CE}(\boldsymbol{F}, \boldsymbol{\theta}_{CE}^{N+1}), \boldsymbol{s}_{true})$

\State $\boldsymbol{\theta} = \boldsymbol{\theta} - \eta\boldsymbol{g}$

\EndFor
\end{algorithmic}
\end{algorithm}

\subsection{TAM-embedded CE}
The support blocks from the same propagation environments are employed as the input of TAM. The generated parameter vector, $\textbf{a}_{task}$, selects important feature vectors in the channel domain. When encountering a new environment, TAM attempts to predict the PDP based on $\mathcal{T}_{support}$, which is the most representative feature of each environment. The attention vector, $\textbf{a}_{task}$, is generated based on the predicted PDP. By performing channel-wise multiplication with the original feature map, the backbone can have a better generalization ability to the new environment.


The TAM plays another essential role in multi-environments adaption. Similar to the self-attention mechanism, the TAM-based method ensures the entire deep learning model to become robust to the data with different distributions. The difference is that self-attention generates weights based on local features (e.g. features of the sample itself), while TAM uses global features (e.g. PDP of the environment) to select important features. The original CE backbone cannot converge when the training dataset contains channel coefficients from different environments. However, with TAM embedded, the system can converge to an optimal minimum with data from all environments.  

It is challenging to analyze in detail how the TAM learns the weight vectors, $\textbf{a}_{task}$, to improve the robustness. But we can try to explain the role of the attention mechanism in an implicit way. For the conventional neural network, all data with different distributions are processed by the same weights of the DNN and there is no dynamic adjustment to adapt to a specific distribution. Such a philosophy of DNN design limits the diversity of trainable data distributions. The operations of TAM are similar to the concept of "divide and conquer", where generating weight vectors for feature maps can be treated as a dynamic adjustment. With such an adjustment, the model can be adapted to the training data with various distributions.

Furthermore, the mechanism of this dynamic adjustment can be learnt by the attention network in the training process of adapting to various environments. Compared with TAM, self-attention has poor performance for blocks in the new environment. Because attention is generated based on the feature map of the block itself, self-attention can only find the basic features with the prior of previous training classes and lacks generalization ability to new environments. In contrast, we use TAM to enhance the generalization ability to the new environment. TAM performs as a meta-learner, which aims to learn to make the backbone generalized to various tasks. After being trained by a large number of pilot blocks from different environments, TAM can find out the environment features based on the experience when it encounters pilot blocks of the new environment. The environment-specific attention helps the backbone to be generalized to the new environment.
\subsection{CAM-based Initialization Network}
One of the limitations of the TAM-embedded CE backbone is that the attention generated by TAM is independent of query blocks $\mathcal{S}_{query}$. In other words, for the same propagation environment, different query instantaneous channels with various feature maps will have the same weights if support blocks $\mathcal{T}_{support}$ are the same, which results in attention maps becoming less efficient. Furthermore, TAM cannot introduce extra channel features in the support blocks to the query blocks since TAM only works on re-weighting existing features from the query blocks. Due to more pilots in each support block, we intend to develop a model to allow $\mathcal{S}_{query}$ to learn additional features, which are contained in those positions without pilots. This learning efficiency should be further improved by exploiting the correlation between support blocks $\mathcal{T}_{support}$ and query blocks $\mathcal{S}_{query}$.

As shown in Figure~\ref{fig:end2end_attention}, we use a CAM-based initialization network to realize that goal. Both $\mathcal{T}_{support}$ and $\mathcal{S}_{query}$ are sent into a pre-trained feature extractor. Then feature maps are used as input of CAM to generate weighted feature maps in the spatial domain. Each pair of feature maps, $\textbf{P}_i$ and $\textbf{Q}_i \in \mathbb{R}^{w \times c}$, are concatenated to form a high-dimensional feature map $\textbf{F}_{initial}$. There are $n$ pairs of $\textbf{P}_i$ and $\textbf{Q}_i$ in total, where $n$ refers to the number of support blocks as mentioned before. The new feature map, $\textbf{F}_{initial}\in \mathbb{R}^{2n\times w \times c}$, is processed by the CNN. With the convolution operation, the channel features at the positions without any pilots in the query block can be learned with the help of support blocks in $\mathcal{T}_{support}$. The output of the CNN is treated as the initialized query blocks, $\textbf{s}_{initial}$, and sent into the CE backbone for further processing.

The function of the CAM-based initialization network is to use $\mathcal{T}_{support}$ to guide CE for $\mathcal{S}_{query}$. CAM focuses on the correlation established by cross-attention so that query blocks with less channel information can learn the rich channel information contained in support blocks more efficiently. The meta-learner in CAM contributes to the new environment adaption and its working mechanism is the same as TAM mentioned above. Through dynamic adjustment in the channel domain based on the global feature and information enhancement in the spatial domain based on each pilot block, a better estimation accuracy can be achieved with the cooperation of the CAM initialization network and the TAM.


\section{Experiments}

In this section, the proposed attention-based CE system is trained and tested under different propagation scenarios. We first introduce the WINNER channel model employed for simulating the propagation effect in different environments. Then we describe the details of the experiment settings. We implement the SwitchNet \cite{jiang2021ai} and test it using the same training and testing set to compare with our proposed attention-based method. We show that the attention-based CE system outperforms SwitchNet. Furthermore, we explore how the CAM-based initialization network and cross-attention mechanism help improve the testing performance. The testing accuracy boundary of $\frac{w}{4}$-pilot case is also considered, where each testing environment has sufficient training data for the CE backbone to learn from instead of only few shots available. With the help of the $w$-pilot block in $\mathcal{T}_{support}$, our proposed method is able to get closer to the testing accuracy boundary or even exceed it. 

\subsection{WINNER Channel Model}

All instantaneous channel coefficients in this experiment are generated by the WINNER channel model (WCM) \cite{bultitude20074}, which has been adapted to various mobile communication scenarios from a local area to a wide area. WCM uses spatial and temporal parameters obtained from the measured CIR to characterize different environments. The measured CIR for each propagation environment is analyzed and processed to get the environment-specific parameters \cite{bultitude20074}, which can be used to simulate the propagation effect for the specific environment. There are twelve different propagation scenarios\footnote{One propagation scenario contains multiple different propagation environments. One PDP represents one environment. By changing the positions of transceivers and propagation conditions in the same scenario, we can obtain various environments.} that WCM can emulate and we choose five of them as the training set while another two are used for testing. The PDP varies according to different environments, which leads to different lengths of instantaneous channel coefficients. In order to facilitate the subsequent training of the DL-based CE model, we unify the number of all channel coefficients to 72 and use zero padding for those channels shorter than 72. Since channel coefficients are complex numbers, we split each one into two real numbers. Therefore, the number of real coefficients for each channel is $72\times 2$.
\begin{figure}[!htb]

\centering
{
\includegraphics[width=0.45\textwidth]{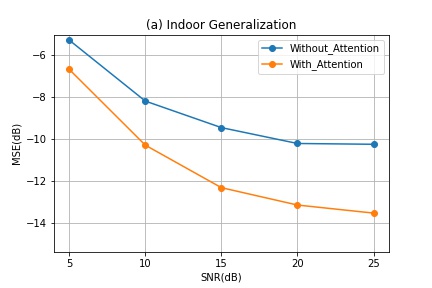}
\label{fig:generalization1}}\\
{
\includegraphics[width=0.45\textwidth]{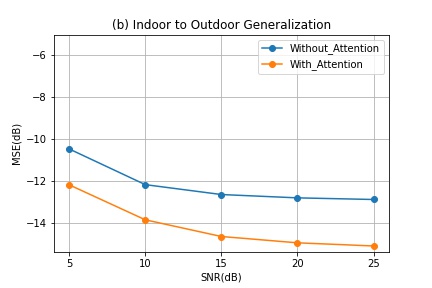}
\label{fig:generalization2}}
\caption{The MSE between true and estimated channel for known environments}\label{fig:generalization}
\end{figure}

\begin{figure*}[h]
\centering
\subfloat{
\includegraphics[width=0.45\textwidth]{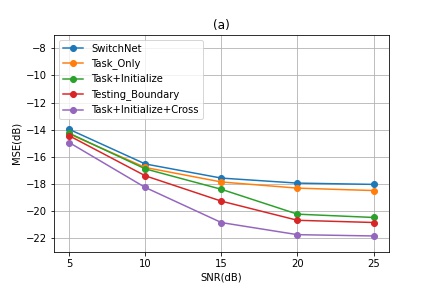}
\label{fig:subfig1}}
\subfloat{
\includegraphics[width=0.45\textwidth]{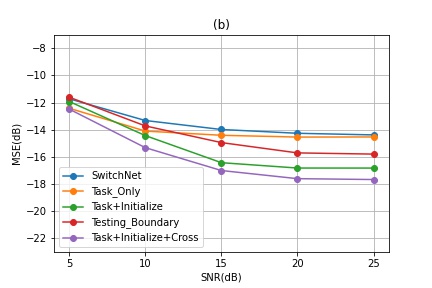}
\label{fig:subfig2}}
\qquad
\subfloat{
\includegraphics[width=0.45\textwidth]{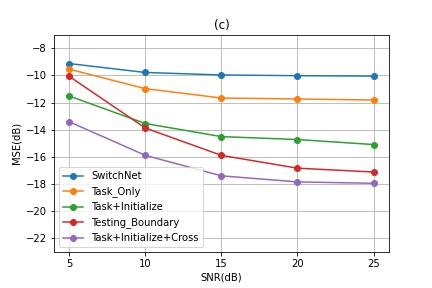}
\label{fig:subfig3}}
\subfloat{
\includegraphics[width=0.45\textwidth]{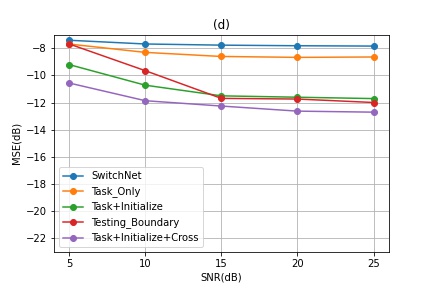}
\label{fig:subfig4}}
\caption{MSE between true and estimated channels for rural macro-cell at different positions (a) (b), and for rural moving network at different positions (c) (d).}
\label{fig:first_two}
\end{figure*}

\begin{figure}[h!]
\centering
\includegraphics[width=0.45\textwidth]{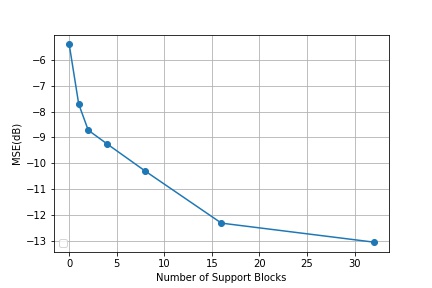}
\caption{The MSE between true and estimated rural channel versus the number of support blocks.}
\label{fig:Number_few_shot}
\end{figure}

\subsection{Experiment Settings}

The training set for five different propagation scenarios are indoor office, indoor-to-outdoor, indoor hotspot, outdoor-to-indoor, and urban macro-cell. It is evident that the average power of each channel tape varies for different environments or the same environment but with different transmitters and receivers positions. We generate 500 PDPs for each propagation scenario for the training dataset by changing the propagation conditions and user positions. Each PDP contains 500 different instantaneous channel coefficients. A significant number of PDPs are required for training the meta-learner in TAM and CAM to learn as many environmental features as possible. When new environmental samples appear, the meta-learner can learn new features by combining the learned features, which guarantees the new task adaption of the backbone.

The testing set includes channels in rural macro-cell and moving networks. Each propagation scenario has five PDPs. The channel output can be formulated as $\bf{y} = \bf{h} \circledast \bf{x} + \bf{n}$, where $\bf{n}$ and $\bf{x}$ are channel noise and channel input, respectively, $\circledast$ refers to the linear convolution, and $\bf{h}$ denotes the CIR. The input of the attention-based CE system is the query blocks with $\frac{w}{4}$ pilots and the support blocks contain $w$ pilots. The number of support blocks required for the new environment \footnote{All new environments are from the propagation scenario that is not included in the training set} will be explored in the following part. All experiment results are obtained by taking the same 500 samples of each environment for testing and then averaging.

\subsection{Baseline}

The proposed attention-based CE system is compared with two baselines, SwitchNet and testing accuracy boundary. For SwitchNet, $\frac{w}{4}$ pilots are employed as the input for the LS CE block, which has the same number of pilots as the CE backbone. The testing environments described in \cite{jiang2021ai} are limited for the SwitchNet, while part of the online testing is implemented with similar propagation conditions but with different max delays. We explore the ability of SwitchNet to adapt to highly separate channels in this part. Furthermore, we will measure the distance between the test performance of the proposed attention-based system and the testing accuracy boundary. We will see whether the testing performance of our proposed method can approach or even exceed the testing accuracy boundary. We also test the effectiveness of each part of the attention-based system. First, we explore how the CE backbone performs with only TAM embedded. We will see, compared with SwitchNet, whether the TAM can help the CE backbone to perform better in the new environment through the dynamic adjustment from the meta-learner. Then the CAM-based initialization network is added to check whether the system can consistently improve its generalization to the new environment. 

\subsection{End-to-End Model Training}
The number of training samples for each environment is $1.25 \times 10^6$ and the training batch size is 128. The whole attention-based CE system is trained in an end-to-end manner. Each training sample includes two categories of data: a frame of $\frac{w}{4}$ pilots block and multiple frames of $w$ pilots blocks. Such multiple frames of $w$-pilot blocks refer to support set $\mathcal{T}_{support}$ in FSL sampling, which is randomly sampled from the training set and belongs to the same average PDP. The loss function and the optimizer used for training are binary cross-entropy and Adam. The model is trained with SNR = 20 dB, while the value of SNR varies from 5 dB to 25 dB during testing.

\subsection{Experimental Results}
Figure~\ref{fig:generalization} demonstrates how the attention mechanism can help the model to improve the CE accuracy in joint training of multiple environments. As mentioned in the previous section, the TAM-based method ensures the DL model to become robust to the data with different distributions. We select two environments from the training set and test the generalization ability under the condition of whether the TAM is applied for CE. Figure~\ref{fig:generalization} shows that the TAM can enhance the adaption for multiple environments, especially in high SNR scenarios.    

Figure~\ref{fig:Number_few_shot} shows the testing accuracy for the rural macro-cell channel versus the number of support blocks. From Figure~\ref{fig:Number_few_shot}, when there is only one block in $\mathcal{T}_{support}$, the performance improvement of CE is the most significant compared with no block since it is easy to recognize some basic channel features from the one-shot pilot block, such as channel taps with high power in the specific environment. As the number of support blocks increases, the performance improves since more implicit features are learnt. The elbow point appears when the number is 16. After the number of blocks exceeds 16, the accuracy improvement is very limited by continuing to increase the blocks. Therefore, the subsequent experimental results are all tested under the condition that the number of blocks contained in $\mathcal{T}_{support}$ is 16. 

During training, TAM allows the CE backbone to be adapted to data in various distributions. Without the TAM, the model has a bad generalization performance using training samples from different environments. Figure~\ref{fig:first_two} compares mean-squared error (MSE) for different design methods. From figure~\ref{fig:first_two}, the model's generalization ability to the new distributed data is enhanced using the task-attention mechanism. The TAM can help the CE backbone to achieve a lower MSE for the new environments compared with SwitchNet. The degree of freedom for the dynamic adjustment by TAM is much higher than SwitchNet. The attention network generates multiple parameter vectors while SwitchNet can only rely on the five parameters for the new environment adaption.

Furthermore, SwitchNet is trained separately for the datasets with different distributions while the attention-based approach uses joint training of these datasets, which allows the model to learn additional information in joint learning from different environments, such as high-level features for the meta-learner training. Therefore attention-based mechanism outperforms SwitchNet in generalization to the new environment. In addition, SwitchNet requires online training steps to be adaptive to the new environment. Our proposed attention-based method is directly generalized to the new environment without fine-tuning steps for online adaption.

\begin{figure}[!htb]

\centering
{
\includegraphics[width=0.45\textwidth]{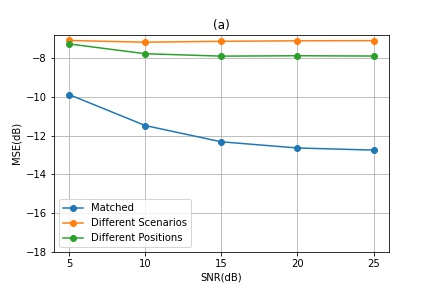}
\label{fig:mismatch1}}\\
{
\includegraphics[width=0.45\textwidth]{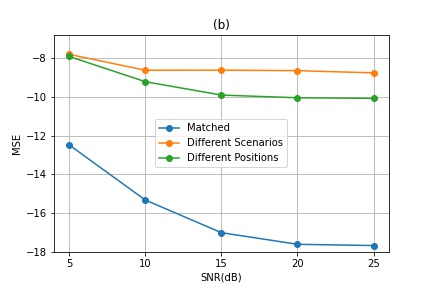}
\label{fig:mismatch2}}
\caption{MSE between true and estimated channels in new environments under match and mismatch cases for rural macro-cell (a) and rural moving network (b)}\label{fig:mismatch}
\end{figure}
The CAM-based initialization network positively affects the TAM-embedded model's generalization to the new environment from Figure~\ref{fig:first_two}. Additional channel information in support blocks can be effectively learned by the query block in the initialization network. Furthermore, We demonstrate that CAM can enhance the system's testing performance. In Figure~\ref{fig:first_two}, the FSL model without CAM-embedded is also considered and it has worse performance compared with the CAM-embedded case. The CAM provides attention in different dimensions from TAM and it is proved in \cite{woo2018cbam} that combining channel and spatial attention leads to consistent improvements for CNN-based models. Figure~\ref{fig:first_two} shows that with the initialization network embedded, the CE system becomes more robust for low SNR. For the high SNR scenario, the initialization network can improve the testing accuracy and is close to the boundary in most cases. With the help of CAM, our proposed model can exceed the $\frac{w}{4}$-pilots testing boundary.

We should emphasize that such MSE performance with support blocks is not due to the high similarity between the test environment and environments in the training sets since we carefully set up the experiments to avoid such circumstances. We employed CNN to do classification for all training and testing channel coefficients, whose label is the environment to which the channel belongs. The classification accuracy is over 95 \%, which indicates that these environments are highly separated and have features that distinguish them from other environments. In addition, we prove that if query blocks $\mathcal{S}_{query}$ and support blocks $\mathcal{T}_{support}$ belong to different environments, which is called 'mismatch'. Two mismatch cases are considered. One case is that these two different environments are obtained from different scenarios. The other is in the same propagation scenario but with different transceiver positions. Both cases lead to varying PDPs for $\mathcal{S}_{query}$ and $\mathcal{T}_{support}$ and result in significant degrading for the testing accuracy. Only the features of the same PDP can give the most accurate CE in this environment, indicating that environmental feature similarities between different PDPs are not high.

Figure~\ref{fig:mismatch} demonstrates that mismatch leads to a significant degrading of the testing accuracy. For the same propagation scenarios, PDPs in different positions share more similar features compared with different scenarios. Therefore, the mismatch for different positions has a better performance than in different scenarios. However, since most channel features are different, such as delay profiles and the number of taps, the gap between mismatch and match performance is enormous.     

\section{Conclusions and Further Directions}
We have realized FSL for new unseen propagation environments in the DL-based data-driven CE model by exploiting an attention-based mechanism. With few pilot blocks sampled from the new environment, the global features of the new environment and correlations between the support and the query blocks can be quickly extracted. Environment-specific and block-specific attention is generated to allow the model to be fast-adapted to new environments. The proposed mechanism outperforms the existing FSL method in the data-driven scenario. 

In this article, we have used CE as an example to realize our novel idea: learn to adapt to new environments using past experience. The same spirit can be utilized in a large group of communication systems and networks, such as end-to-end systems and signal detection, localization, and resource allocation.

For future research, it is desired to investigate the model-driven DL-based wireless communication model for FSL. The model-driven DL-based approaches combine communication domain knowledge with DL models \cite{he2019model}. Compared with the data-driven method, the model-driven method only contains a small number of parameters that need to be trained, which means the demand for the amount of training data is not so significant. The challenge is to find the most critical parameters affecting performance in the specific environment and figure out which environment features are the most closely associated with these parameters. Therefore, the model-driven approach has great potential to deal with FSL problems.


Another interesting research topic is to apply the graph neural network (GNN) to deal with FSL problems in wireless communications. GNN can be designed to capture the dependence of graphs through information transformation between nodes in the graph. Its working mechanism is equivalent to distributed optimization algorithms. GNN has the potential to perform FSL due to its fewer parameters and high computation efficiency owing to its distributed structure. Furthermore, we can employ GNN to enhance the model-driven method, such as the belief propagation (BP) algorithm. GNN can be used to construct the factor graph and extract features from each iteration in parameter updating through the algorithm so that optimum parameters can be learned.
\bibliographystyle{IEEEtran}
\bibliography{reference}

\end{document}